\documentstyle[12pt]{article}
\begin{document}
\begin{titlepage}\begin{center}
\vskip 0.4in
{\Large\bf Electron-Neutrino Degeneracy and Primordial Nucleosynthesis\footnote{presented at the Seventh Asian-Pacific Regional Meeting of the IAU, Aug. 19-23, 1996, Pusan, Korea}}
\vskip 0.8in
{\large  Jong Bock Kim, Joon Ha Kim, and Hyun Kyu Lee}\\
\vskip 0.1in
{\large \it Department of Physics, Hanyang University,} \\
{\large \it Seoul 133-791, Korea}\\
\vskip .4in

{\bf ABSTRACT}\\ \vskip 0.1in
\begin{quotation}
\noindent
We discuss the possible
ranges of electron neutrino degeneracy which is consistent with the inferred
primordial abundances of the
light elements. It is found that the electron neutrino
degeneracy, $ |\xi_{e}|$,
up to order of
 $ 10^{-1}$ is consistent with the present data.
\end{quotation}
\end{center}
\end{titlepage}

\newpage
The big bang nucleosynthesis(BBN)
has been considered as one of the important cosmological processes which
provides the valuable limits on  the number of  cosmological
and particle physics parameters, for example, the number of species  of light
 neutrinos, and
the baryon number density of our Universe.

However, given the uncertainties of the inferred abundances, the possibility
of variations from the standard BBN(SBBN) has been studied in various
contexts(Maleney93).  Moreover, recent measurements of deuterium(Rugers95,
Tytler96) abundance,
new estimations of $ ^4He$ (Izotov96) together with the  recent  statistical
assesment of the SBBN(Hata95, Copi95) encourage the detailed study of the possible
variations. In SBBN the neutrinos are assumed to
be  very  simple object although only little of neutrino properties  are
known. For example,
mass, magnetic moment, oscillation, and neutrino
degeneracy are among them , which should be studied
in deatil to see how the BBN constraints them and how those properties affect the
primordial nucleosynthesis.

The concordance between the  predictions and the observations of
 the light element
 abundances seems to
be remarkable with $\eta \sim 3$, which lead to the nonbaryonic-matter dominated Universe.
By introducing the neutrino degeneracy  it is possible to reduce nonbaryonic matter
considerably(Kang92) since it allows more baryonic matter which participate
 into nucleosynthesis.
 It is also suggested(Kim95) that if  primordial abundance of $^{11} B$ can be
measured it might be a possible
test,  since it depends quite sensitively on the baryon number density.

In this work, we estimate the possible ranges of electron
neutrino degenercay using the sets of recently observed(inferred)
abundance data of light elements .

The evolution of the Universe is described by the expansion rate $H$,
$H^{2} = \frac{8\pi}{3} G \rho_{rad} \label{hubble}$
,where $G$ is the gravitational constant.
$\rho_{rad}$ is the energy density of the light particles
which  includes the contribution from the possible nonzero neutrino
degeneracies.
While the contribution of
neutrino degeneracy($\xi_e = \mu_{\nu}/T$ for the
degenarate neutrino with chemical
potential $\mu_{\nu}$)  speeds up the expansion rate  of the Universe,
the  degeneracy
of electron neutrino, $\xi_{e}$ also affects the weak interaction rates, which
together with
the expansion rate determine the weak interaction freezing temperature, $T_{f}$,
and  it
 controles the neutron to proton ratio at the beginning of nucleosynthesis.
The effects of muon- and tau- neutrino degeneracy are only on the expansion rate
which can be
possibly accommodated by modifying the gravitational constant $G$(Kim95).
In this work,  we take $G$ as  the present observed value of gravitational
 constant and assume no degeneracy of muon- and tau- neutrinos to see more transparently the
  role of
the electron neutrino degeneracy.  The three light neutrinos are considered
to be effectively massless during the nuclear synthesis  and
no neutrino oscillation is assumed.

Using the extended reaction network(Kim95) with neutron mean life
time taken to be 887.0$\pm 2.0 sec$, we calculate the abundances as a function
 of $\xi_{e}$ and
the baryon to photon ratio $\eta$ to find out the possible ranges which are consistent with
the inferred abundances of light elements.

\noindent  For deuterium abundance, we take both high and
low values from QSO observations. The low abundance of deuterium, $[D/H]
 = (1.7 - 3.5) \times
 10^{-5}$ which is similar to the ISM value ,
 has been estimated by (Tytler96), while higher abundance, $[D/H] = (1.5 - 2.3)
  \times
 10^{-4}$by (Rugers95).  We use  $0.226 \leq Y_p \leq 0.242$ for $^4 He$ and
  for $^7
 Li$, $0.7 \leq [^7 Li/H] \times 10^{-10} \leq 3.8$(Olive95).

\noindent The allowed regions for $\xi_e$ and $\eta$ are serched in $\xi_e$ - $\eta$ plane
which are consistent with these  data.

With low value of $[D/H]$, the case is similar to the anaysis  which
leads to the recent `crisis' (Hata95) where the  possible variations of SBBN
 , for example , smaller number of  of light
neutrino species, nuetrino degeneracy or underestimation of $^4 He$ abundance
have been suggested.
By allowing the neutrino asymmetry , we can
 have good fits to the
inferred abundances. The allowed ranges for  $\xi_e$ and $\eta$  are
$-0.01 \leq \xi_e\leq 0.08$ and $4 \leq \eta \leq 6$, respectively .
Replacing the deuterium abundance  with higher value , the permitted range for
the $\eta$ becomes  smaller
 $ 1.4 \leq \eta \leq 2.0$.
The permitted range for positive
electron neutrino degeneracy is very limited because the adopted
 deuterium abundance prefer the
lower $^4 He$ abundance. The permitted range is found to be $-0.07 \leq \xi_{e}
\leq 0.02$. In connection with the two distinct deutrerium
abundance estimations, it is observed that the positive electron neutrino
 degeneracy is preferred for lower value
of deuterium abundance while the larger range of negative $\xi_e$ is possible
 for the higher deuterium abundance.

If we adopt  the high  abundance of $^4 He$ by Izotov et al.(Izotov96),
$0.240 \leq Y_p \leq 0.246$ up to 1-$\sigma$ statistical error,
 the permitted ranges of $\xi_e$ and $\eta$ are  changed only slightly.
For example, we can observe
the permitted range of $\xi_e$ is a little bit enlarged  in
negative direction, up to -0.09.

From this analysis, we can observe that the nonvanishing electron neutrino
 degeneracy is
allowed up to $|\xi_e | \sim 0.1$. It is interesting to note that this value
 is not inconsistent
with the recently proposed limit from the neutrino oscillation
consideration(Foot95) and the
large scale stucture analysis(Madsen95).

  For the case of  low value of $D/H$,  we can see that the permitted ranges of
   $\eta$ with nonvanishing $\xi_e$ are higher than SBBN case. Then the
   observation
   of the heavier elements $^9 Be$ and $ ^{11} B$ might be a good test(Kim95)
   of nonvanishing electron neutrino degeneracy  aginst SBBN,
   since they are very sensitive on $\eta$ in the  range of our interest.
However
 if the estimation by Izotov et al.(Izotov96) can be adopted  then there is
 essentially no differences in the permitted ranges of $\eta$  and
the abundances of $^9 Be$ and $ ^{11} B$,it might not be a best
way of
 testing the  neutrino degeneracy. Also for the larger deuterium abundances
 the permitted range of $\eta$ is not sensitive to the neutrino degeneracy.
   However, since $^9 Be$ and $ ^{11} B$ are
sensitive on $\eta$ in the range of our interst,  the measurements of those
abundances themselves
are still needed in connection with the current issues in deuterium abundances
which is also very sensitive on $\eta$.

In summary,  while the issue whether the implication of the nonvanishing
electron neutrino
degeneracy can be quite different from SBBN so as to be tested, for example
in  the primordial abundances of heavier element like
$^9 Be$ and $ ^{11} B$, needs more  observational data and more
detailed analysis of the  data,
we observe that
the nonvanishing electron degeneracy, up to $|\xi_e| \sim 0.1$, can be a
possible
variation of SBBN.

HKL thanks Institute of Nuclear Theory at the University of Washington for its
hospitality where a part of this work is done while he is
participating  INT 96-2
workshop. He also thanks George Fuller and Gary Steigman for useful
discussions.  This work was supported by the Ministry of
Education(BSRI-95-2441) and in part by the Korean Science  and Engineering
Foundation and also in part by the Hanyang University Research Grant(1996).

\newpage

\noindent {\bf References}\\

\noindent C. Copi et al., Phys. Rev. Lett. {\bf 75},
3981(1995)\\

\noindent R. Foot and R. R. Volkas, Phys. Rev. Lett. {\bf 75},
4350(1995)\\

\noindent N. Hata et al., Phys. Rev. Lett. {\bf 75},
3977(1995)\\

\noindent Y.I. Izotov, T.X. Thuan, and V.A. Lipovetsky,
submitted to ApJ, 1996.\\

\noindent H. Kang and G. Steigman, Nucl. Phys. {\bf B372},
494(1992)\\

\noindent J. B. Kim and H. K. Lee, Astrophys. J.
  {\bf 448}, 510(1995); J. B. Kim, and H. K. Lee, J. Korean Phys. Soc.
{\bf 28}, 662(1995)\\

\noindent G.B. Larsen and J. Madsen, Phys. Rev. {\bf D52},
4282(1995)\\

\noindent R.A. Maleney and G. Mathews, Phys. Rep. {\bf 229},
147(1993)\\

\noindent K. Olive and G. Steigman, ApJ. Suppl. {\bf 97},
49(1995)\\

\noindent M. Rugers and C. Hogan, ApJ. {\bf 459}, L1(1995)\\

\noindent D. Tytler, X. Fan, and S. Burles, Nature {\bf 381},
207(1996)\\

\end{document}